# Bimodality in Low Luminosity E and S0 Galaxies

A.E. Sansom[1*] and M.S. Northeast[1]
[1]*Centre for Astrophysics, University of Central Lancashire, Preston, Lancs PR1 HE*

**ABSTRACT**
Stellar population characteristics are presented for a sample of low luminosity early-type galaxies (LLEs) in order to compare them with their more luminous counterparts. Long-slit spectra of a sample of 10 LLEs were taken with the ESO New Technology Telescope, selected for their low luminosities. Line strengths were measured on the Lick standard system. Lick indices for these LLEs were correlated with velocity dispersion ($\sigma$), alongside published data for a variety of Hubble types. The LLEs were found to fall below an extrapolation of the correlation for luminous ellipticals and were consistent with the locations of spiral bulges in plots of line strengths versus $\sigma$. Luminosity weighted average ages, metallicities and abundance ratios were estimated from chi-squared fitting of 19 Lick indices to predictions from simple stellar population models. The LLEs appear younger than luminous ellipticals and of comparable ages to spiral bulges. These LLEs show a bimodal metallicity distribution, consisting of a low metallicity group (possibly misclassified dwarf spheroidal galaxies) and a high metallicity group (similar to spiral bulges). Finally, they have low $\alpha$-element to iron-peak abundance ratios indicative of slow, extended star formation.

**Key words:** galaxies: abundances – galaxies: evolution – galaxies: elliptical and lenticular, cD

# 1 INTRODUCTION

Bender, Burstein & Faber (1992) suggested that dynamically hot galaxies (i.e. high mass, compact spheroidals such as spiral bulges and giant Es) form increasingly as a result of mergers of whole galaxies as opposed to gaseous mergers of smaller systems. That is to say, the more massive a galaxy, the more strongly a hierarchical merger mechanism may have affected the star formation history (SFH) of the galaxy. If this is so, then less massive, low luminosity galaxies could form important building blocks to larger E galaxies. Recent ideas about the build up of galaxies emphasise the fact that more massive galaxies may have formed from mergers of less gas-rich galaxies, and as such may not be accompanied by merger induced star formation, although this is a controversial subject area (e.g. van Dokkum 2005; Pozzetti et al. 2007; Donovan et al. 2007). The role of low luminosity galaxies in building the Hubble sequence is still unclear. This paper looks at the observational properties of a sample of low luminosity elliptical and lenticular galaxies

* E-mail: aesansom@uclan.ac.uk

(here collectively referred to as LLEs) in contrast with more luminous galaxies to probe their relative star formation histories through simple stellar population (SSP) fitting to observed spectral line strengths. The LLEs discussed here are more luminous than dwarf spheroidal (dSph) galaxies and their properties are also compared to dSphs in this work.

Some previous studies involving LLEs have considered their stellar velocity dispersion ($\sigma$) in relation to the line strengths, as well as sought to identify ages (e.g. Halliday, PhD thesis, 1998, Caldwell, Rose & Concannon 2003 – hereafter CRC03). However, most of these studies suffered from relatively large errors. In Proctor & Sansom 2002 (hereafter PS02) correlations between line strengths and $\sigma$ were compared for different Hubble types. Steep relations were found for metal sensitive line strength versus $\sigma$ for spiral bulges, but those data set lacked early-type (E and S0) galaxies with low $\sigma$ values, below about 80 kms$^{-1}$. Therefore a direct comparison between bulges and ellipticals of similar $\sigma$ could not be made. Whether the correlations found for spiral bulges exist at low $\sigma$ for E/S0 galaxies is investigated in this paper. The trends taken from Kuntschner 2000 (hereafter K00) for more luminous early-type galaxies showed shallower slopes than observed for spiral bulges in plots of index versus $\sigma$. It is important to see how LLEs relate to this, to test if LLEs are like spiral bulges at similar $\sigma$ or not. The principal aim of the present work is therefore to investigate:-

- How will observed low $\sigma$ galaxies compare between early and late-type galaxies?

- Will low $\sigma$ E/S0 galaxies fall on extrapolations of the line strength versus $\sigma$ trends for luminous ellipticals?

- From derived parameters, what average ages and abundances do the LLEs have?

- Hence what galaxy formation mechanism do the observations of LLE galaxies appear to support?

LLEs are expected to be generally of lower $\sigma$ than more luminous E galaxies, from the mass luminosity relation and the Virial theorem. As such LLEs are a good choice for low $\sigma$ targets. A few of the targets selected were also of known low $\sigma$. In this paper, analysis of measurements of the kinematics and line strengths of a sample of LLEs is carried out; correlations are compared to trends reported in PS02 and elsewhere for different Hubble types, in order to compare their structure, ages and abundances at fixed $\sigma$. The observations of LLEs discussed in this paper were aimed at providing sufficient signal-to-noise (S/N) in the central regions of these galaxies to address these questions.

In Sections 2 and 3 the observations and initial data reductions are described. Section 4 presents derivations of galaxy kinematics, including calibration to the common Lick scale (e.g. Worthey et al. 1994) described in an Appendix. In Section 5, extraction of Lick Indices and comparisons with other Hubble types are described. In Section 6 measurements are made of derived parameters and key results on ages and abundances are presented. Section 7 summarises the main conclusions.

# 2 OBSERVATIONS

Galaxies were selected as being LLEs of low or else unknown σ, with distance (≤ 20 Mpc), and B band Magnitude ($M_B > -18$), with the exception of NGC 2784 ($M_B$=-19.15), which was used as a brighter control. Information on these galaxies is catalogued in "The Third Reference Catalogue of Bright Galaxies" (de Vaucouleurs et al. 1991), hereafter RC3. Most of the galaxies are in groups or in the Fornax cluster (detailed in Table 1). Observations were made with the European Southern Observatory's (ESO's) New Technology Telescope (NTT), with the ESO Multi-Mode Instrument (EMMI) used in double spectrograph mode, on the nights of the 25 to 27 November 2003.

The wavelength range was chosen to include the same Lick indices as in PS02 to enable direct comparison with those data. Observations in EMMI are done in REMD (Red Medium Dispersion Spectroscopy mode) for λ > 400 nm. The red arm of the EMMI instrument was chosen, in order to cover more Lick indices simultaneously. EMMI has a pixel scale of 0.166" per pixel. The red CCD has 4152 x 4110 pixels, spread equally over four chips. The resolving power achievable, using grating 7, is 2600. At a central wavelength of about 500 nm, this gives a spectral resolution of 0.192 nm FWHM. For the unvignetted wavelength range used (405 to 560 nm across 3754 pixels) a dispersion of 0.041 nm per detector pixel was obtained. A slit was selected of dimensions 1" by 300" (in the wavelength and spatial directions respectively) to detect light from across a galaxy and from neighbouring sky.

Twelve galaxies were observed, with the slit running along the minor axis of each spheroid in turn. The minor axis was chosen for direct comparison with results for spiral bulges and ellipticals from PS02. The position angle of NGC 1374 is nearly zero and not significantly out of alignment with the minor axis of the neighbouring low luminosity elliptical NGC 1375; therefore a PA of 0 degrees was used to observe both galaxies in that case. NGC 1374 acts as a second control, at intermediate velocity dispersion. Between two and six exposures of approximately twenty minutes each were made for each galaxy, dependant upon the brightness of the galaxy and observational conditions.

Three flux and 17 Lick calibration stars (from Worthey et al. 1994) were observed. A few of the observed galaxies have spectra with strong emission lines. One emission line example is ESO 118-34, which has an HII region-like emission spectrum (Buson et al. 1993). Relevant characteristics of galaxies and stars are noted in Table 1 and Appendix Table A1 respectively. Calibration images (such as biases and flat-fields) were generated during the day, or else near twilight along with the star exposures. Several calibration arc images were taken at various instrument rotation angles since the wavelength scale is known to shift across the detector, for different rotation angles. There were some minor instrumental difficulties with calibration lamps, which were solved during the observing run. Heavy cloud cover over the first night (with humidity near to the telescope dew point), and partial cloud cover over the second, resulted in lower photon-counting statistics in some exposures (particularly NGC 59 and NGC 3125).

| Galaxy | Type | T | $B_T^o$ | $M_B$ | $r_e$ ('') | Environment | Exp (sec) | Air Mass Range |
|--------|------|---|---------|-------|-----------|-------------|-----------|----------------|
| (1) | (2) | (3) | (4) | (5) | (6) | (7) | (8) | (9) |
| NGC 59 | .LAT-* | -2.5 | 13.05 | -15.57 | 21.73 | Sculptor group NOG=11 13 gals | 7200 | 1.012-1.074 |
| ESO 358-59 | .LX-.. | -3.0 | 13.98 | -16.25 | 12.51 | Fornax cluster NOG=208 54 gals | 7200 | 1.009-1.036 |
| NGC 1375 | .LX.0*/ | -2.0 | 13.11 | -16.34 | 16.49 | Fornax cluster NOG=208 54 gals | 7200 | 1.008-1.054 |
| NGC 1331 | .E.2.*.R | -5.0 | 14.15 | -16.69 | 9.27 | Eridanus group NOG=207 51 gals * | 7200 | 1.052-1.266 |
| ESO 157-30 | .E.4.*. | -5.0 | 14.51 | -16.70 | - | Dorado group NOG=26 16 gals * | 9000 | 1.112-1.209 |
| NGC 1373 | .E+..*. | -4.3 | 14.04 | -17.04 | - | Fornax cluster NOG=208 54 gals | 3816 | 1.035-1.274 |
| ESO 118-34 | .L..OP*s | -2.0 | 13.45 | -17.07 | 6.27 | Group NOG=249 8 gals | 7200 | 1.168-1.287 |
| NGC 3125 | .E...?. | -5.0 | 13.02 | -17.53 | 7.89 | Group NOG=439 9 gals | 2400 | 1.112-1.158 |
| NGC 2328 | E-SO | -2.8 | 12.87 | -17.70 | - | Isolated? (OMS01) | 7200 | 1.028-1.046 |
| NGC 1411 | .LAR-* | -3.0 | 12.12 | -17.94 | 13.71 | Group NOG=214 12 gals | 7200 | 1.036-1.061 |
| NGC 1374 | .E... | -4.5 | 11.91 | -19.10 | 24.38 | Fornax (pair with NGC1375) | 7200 | 1.008-1.054 |
| NGC 2784 | .LASO* | -2.0 | 10.3 | -19.15 | 26.74 | Group (CM04) 5 gals | 2400 | 1.080-1.119 |

**Table 1.** Galaxies observed with the NTT, ordered by absolute B magnitude. Hubble type, T-Type and half light radii ($r_e$) are from RC3. B-band apparent magnitudes (4th column) and absolute magnitudes (5th column) are from HyperLeda. The distance for NGC 59 was obtained from Karachentsev et al. 2004. Missing values are shown as "-". Environments: NOG=Nearby optical group number from Giuricin et al. 2000. The number of galaxies in the group with B<14 mag is given to indicate the size of the group. Some group name are from Garcia 1993 (marked *). The Eridanus group is described in Omar & Dwarakanath 2005. Additional information on environments was obtained from the following references: CM04 = Chamaraux & Masnou 2004; OMS01 = Oosterloo, Morganli & Sadler 2001. Total NTT exposures are given in the 8th column. Air mass range is given for all the exposures taken for each galaxy.

# 3  DATA REDUCTIONS

Basic data reductions were carried out using Starlink and IRAF software packages. Error propagation was processed in parallel with reductions carried out on the data, using variance arrays. FIGARO was used for cosmic ray cleaning, wavelength calibration, sky subtraction and flux calibration. KAPPA was used for image transformation and CCDPACK was used for image mosaicing. Further details of the data reductions can be found in Northeast 2006.

Data from the NTT comes in a 4 frame, separate image FITS format, with one frame for each EMMI CCD chip. The CCD chips were reduced individually. They had different bias levels, both between each other and from within to outside the slit extent across the chips. Characteristics of the EMMI CCD detector were obtained from the ESO web site http://www.ls.eso.org/lasilla/sciops/ntt/emmi/. After debiasing, the error remaining in the images was assumed to be dominated by Poisson statistics and 2-d variance arrays were created from the counts data. Cosmic ray cleaning was carried out on the data files. Master dome flats were made by stacking dome flats for a given night and dividing out the average spectral shape. Calibration and science data were divided directly by the normalized master flatfield for each night.

The rotation angles of each arc per science exposure were investigated, within and between nights. Comparison of the shift in wavelength direction with rotation angle, relative to a reference arc from the first night, revealed a significant variation in shift between nights and at larger relative rotation angles. The choice of arcs for wavelength calibration was therefore made by matching each science exposure to an arc of the nearest rotation angle setting, within a night. A 2-d spectral calibration was carried out and the data were scrunched to a linear wavelength scale. Variance arrays were scrunched in parallel for each science object. The root mean square (RMS) position of the centroid of the sky line (OII forbidden line at 551.7 nm) was investigated across each calibrated image as a wavelength calibration check. Wavelength calibration accuracy was about ±0.019 nm. Multiple exposures of a galaxy were aligned and stacked, finding the median of individual stacked pixels. This helped to remove residual CRs, whilst correctly handling the data and variance arrays.

Sky background was subtracted before extinction corrections and flux calibrations. Uncertainties in the sky background were estimated using different selections for the sky region. All science objects were extinction corrected using values appropriate for the La Silla site. Variance arrays were recalculated by running the extinction correction twice on the original variance arrays. For flux calibrations, interpolated spectra were created that excluded absorption line regions in the flux standard star observations. Flux calibration star spectra were divided into the interpolated published spectrum (tabulated values from http://eso.org/observing/standards/spectra/stanlis.html) following normalization by the mean. From the normalised instrumental response derived using HR 8634 and HR 3454, there was no evidence of any nightly variations. The reproducibility of the shape of the normalized instrumental response was good to a ~2 percent. Flux calibrations of galaxy and Lick star spectra were therefore performed using a mean instrumental response constructed from the normalized response curves for HR 8634 and HR 3454.

The S/N in the galaxy spectra was found to be typically ~75 per pixel around 500 nm. The S/N was somewhat lower in NGC 59 and NGC 3125 due to more cloud cover or lower exposure. The flux calibrated 2D galaxy spectra contained some emission lines. These were generally narrow. The interactive program FIGARO:CLEAN was used to manually remove emission lines (plus any residual CRs) from galaxy frames prior to kinematic measurements. These data were then used for kinematic and index determinations. Lick indices most affected by line emission include Fe5015, and the Balmer lines.

For the galaxies, spectra were extracted by first finding the luminosity weighted centre. Spectra were extracted from the central 3.6" regions of galaxies so as to be consistent with the WHT data set of PS02, for comparison with those data. Galaxy and Lick star spectra were converted to FITS files for use in IRAF. The files were further converted and the wavelength axis of each relevant spectrum was reconstituted by the STSDAS command STRFITS, which uses the start and end wavelength in the header. Measurement of σ and radial velocities could then be made, as discussed in Section 4.

# 4 GALAXY KINEMATICS

Measurements of galaxy kinematics were necessary for two principal reasons. Firstly so that line strength measurements could be corrected for velocity dispersion and, secondly, to investigate possible correlations between line strength and velocity dispersion. Line strength measurements corrected for velocity dispersion broadening could then be used to determine galaxy ages and abundances (as described in Section 6).

Before galaxy kinematics could be measured, radial velocities of the Lick calibration stars were determined using cross-correlation. Barycentric corrections were made for comparison to published values. A template star spectrum was used to set the absolute scale for radial velocities of the Lick stars. The radial velocity of the template star was determined from the mean Doppler shift of atomic lines. Line positions were found using Gaussian fitting, which was accurate to ~4 kms$^{-1}$. Comparison of measured radial velocities to published data produced an overall mean difference of about -4 kms$^{-1}$ with a standard deviation of about 28 km$^{-1}$.

The radial velocities (v) and velocity dispersions ($\sigma$) for the galaxies were simultaneously measured using the STSDAS routine FQUOT, which uses the Fourier quotient technique (Sargent et al. 1977). Results could not be obtained for NGC 1373 or NGC 3125. The data for NGC 3125 turned out to be too low signal-to-noise to obtain kinematics. It was not clear why the template match to NGC 1373 was poor. Table 2 shows that NGC 1375 had a particularly large discrepancy between measured and published $\sigma$ values. The reason for this is unknown. However there is a wide spread of measured $\sigma$ for this galaxy, measured by different researchers (i.e. 56 kms$^{-1}$ by K00 and 80 kms$^{-1}$ by D'Onofrio et al. 1995). This is greater than associated reported errors (i.e. about 10 sigma), indicating additional systematic errors may exist between published results. The $\sigma$ reported in HyperLeda (Paturel et al. 2003) are generally averages, weighted by reported errors and may cover different regions of the galaxy. In Table 2, velocities from redshifts given in SIMBAD and dispersions from HyperLeda are compared where possible to values measured here.

| Galaxy | Observed | | Published | | $\Delta\sigma$ | $\Delta v$ |
|--------|----------|---|-----------|---|-----------|------|
|        | $\sigma$ | v | $\sigma$ | v | | |
| NGC 59 | 37 (2) | 377 (4) | - | 367 (11) | - | -10 |
| ESO 358-59 | 51 (4) | 1030 (14) | 46 (5) | 1015 (26) | -5 | -15 |
| NGC 1375 | 48 (1) | 761 (52) | 69 (2) | 732 (44) | 21 | -29 |
| NGC 1331 | 57 (3) | 1204 (20) | 55 (5) | 1315 (99) | -2 | 111 |
| ESO 157-30 | 56 (9) | 1537 (15) | 44 (11) | 1481 (91) | -12 | -56 |
| ESO 118-34 | 52 (1) | 1118 (3) | - | 1164 (23) | - | 46 |
| NGC 2328 | 33 (5) | 1127 (2) | - | 1159 (-) | - | 32 |
| NGC 1411 | 154 (1) | 1033 (14) | 144 (1) | 994 (32) | -10 | -39 |
| NGC1374 | 160(2) | 1360 (21) | 178 (5) | 1332 (36) | 18 | -28 |
| NGC 2784 | 236 (2) | 704 (3) | 225 (9) | 704 (6) | -11 | 0 |
| | | | | **Mean:** | 0 | 1 |
| | | | | **S.t.d.:** | 14 | 50 |

**Table 2.** Velocity dispersion ($\sigma$) and recession velocity (v) in kms$^{-1}$, as measured for NTT data, compared where possible with values from HyperLeda and SIMBAD respectively. Errors are given in brackets. For measured values, the data and errors are as given by FQUOT. The last two columns represent the difference between published and observed values. Mean difference and standard deviation (S.t.d.) are given on the last row.

The mean differences between published and observed kinematics in Table 2 do not show any obvious systematics. The spectra of individual galaxies were matched by 3 to 8 different star templates by FQUOT. The matching of templates, either G or K types, to galaxy spectra was not found to be preferential to any spectral type. Before measuring indices from the Lick star spectra they were first shifted to rest wavelengths. For all measurements of indices from the NTT data the correction to the Lick spectral resolution (as described in Appendix A) was performed. From the half-light radii and velocity dispersions given in Tables 1 and 2 respectively, the dynamical masses of these LLEs lie in the range $\sim 10^9$ to $2 \times 10^{10}$ $M_\odot$.

## 5 LINE STRENGTHS

The software used to measure line strengths (BANSI) was initially written by Sansom and expanded by Proctor (2002). This software measures pseudo equivalent-widths, as well as handling flux calibrations to Lick, kinematic corrections and errors, making use of the variance arrays. Calibrations to the Lick/IDS system were performed in accordance with the procedures described in PS02.

This section investigates how correlations between line strengths (given in Table 3) and $\sigma$ extend to lower $\sigma$ values. To this end data taken with the William Herschel Telescope (WHT), from PS02, was arranged in plots of line strength versus $\log(\sigma)$ together with the current NTT data. These plots are shown for predominantly metal-sensitive indices versus $\log(\sigma)$ in Figure 1a and Figure 1b and for predominantly age-sensitive indices versus $\log(\sigma)$ in Figure 2.

Confidence in the trend lines for each of the line-strengths versus $\log(\sigma)$ in Figure 1a and Figure 1b and Figure 2 is represented by Pearson's correlation coefficient (r). In Table 4 correlations are given for the early-type galaxies including both the present sample of mainly LLEs and the brighter E/S0 galaxies from PS02. In Table 5 similar correlations are given for the present sample and the spiral bulges from PS02. In Tables 4 and 5 each correlation is deemed significant if the Pearson's r is above a 1% level of significance for a two-tailed test.

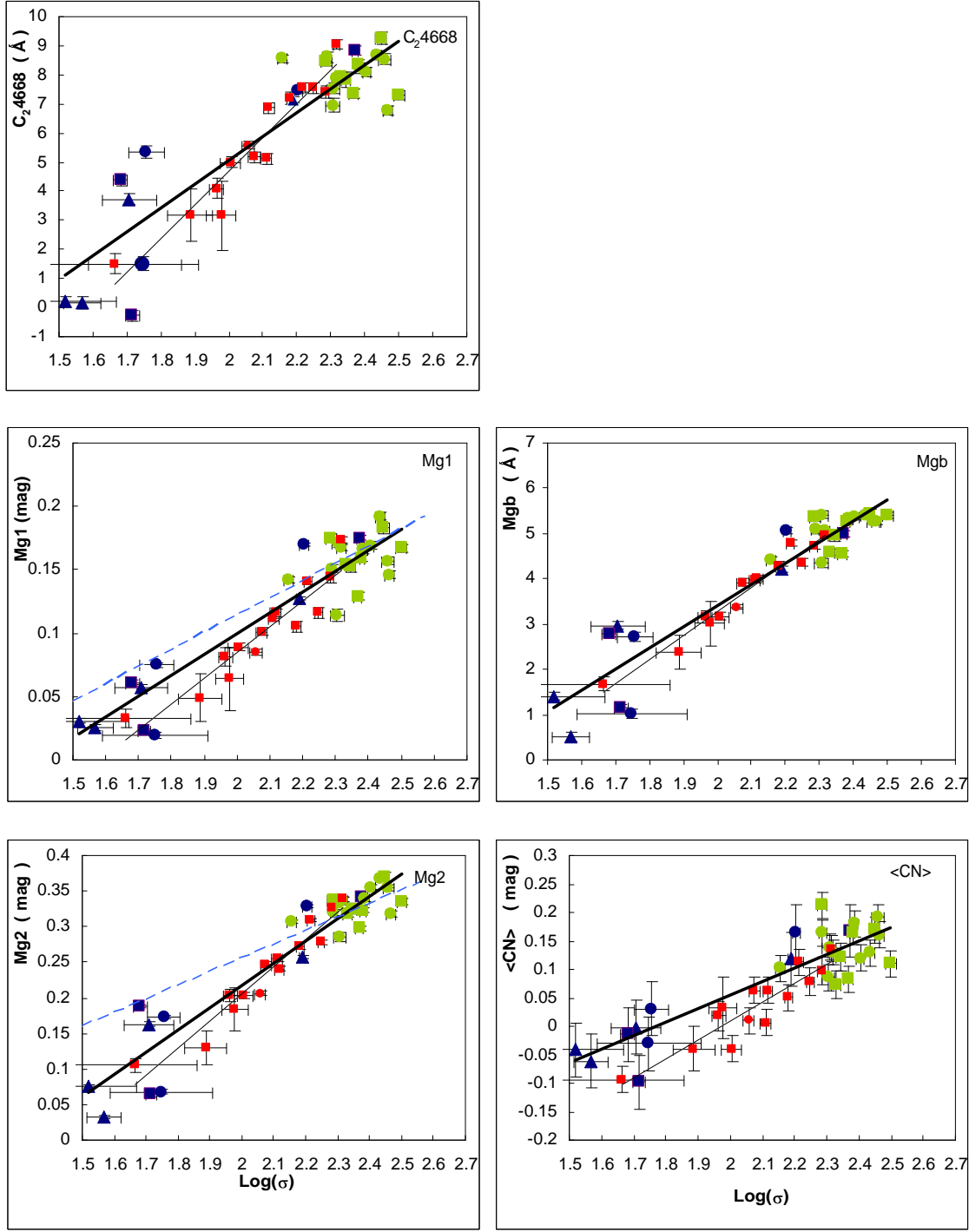

**Figure 1a.** Predominantly metal-sensitive Lick indices plotted against log(σ), for magnesium and carbon sensitive features. The units of σ are kms⁻¹; those of the indices are given on the plots. <CN> is the average of $CN_1$ and $CN_2$. Thin black lines are fits to the spiral bulges (PS02). Thick black lines are fits to the early-type galaxies (PS02 + current data). Dark blue points (circle for E, square for S0, triangle for E-S0) are from the current NTT data. Light green points are early-type galaxies (E and S0) from PS02. Red points are late-type galaxies (S0/a and spirals) from PS02. The dashed blue lines are for early-types from K00, extended from their original range to low velocity dispersions for comparison to our data. In the top right hand corner of each plot, the relevant Lick Index is noted.

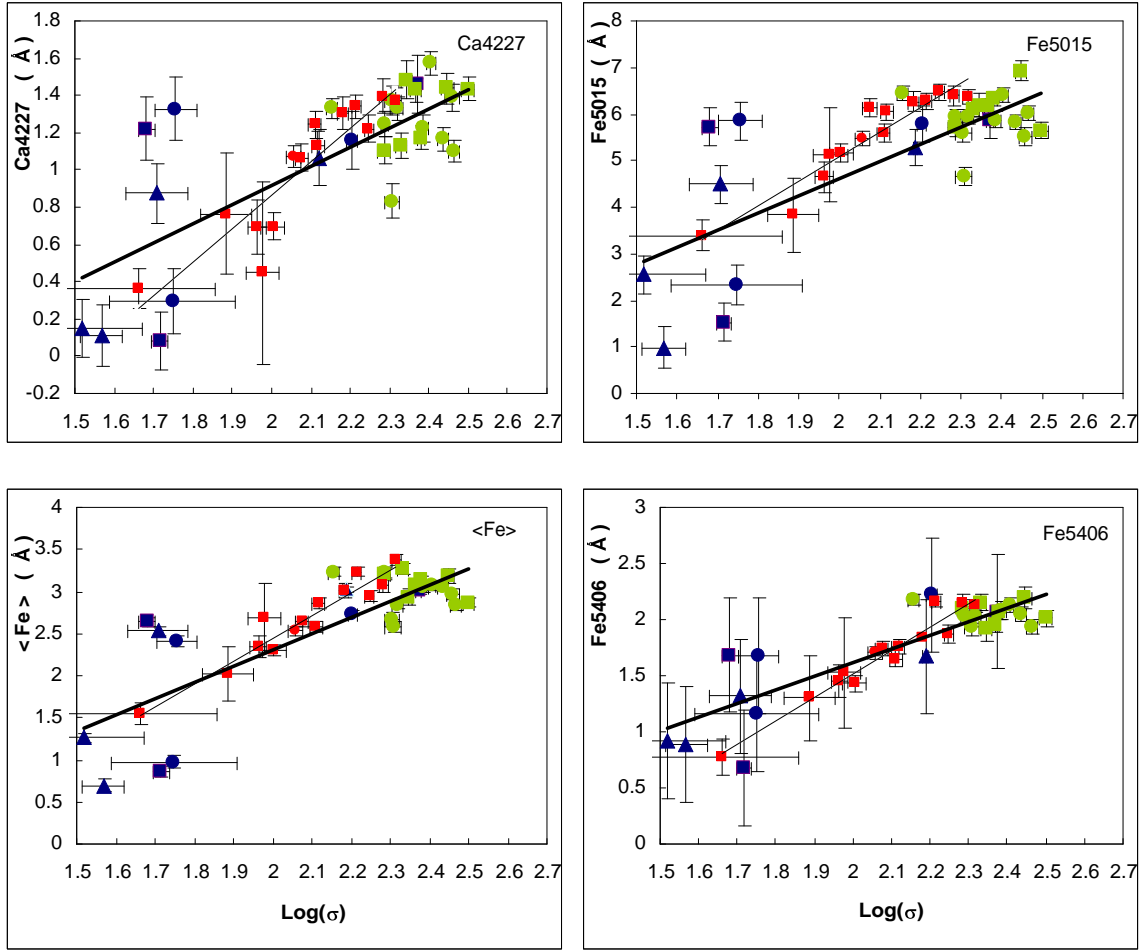

**Figure 1b.** Predominantly metal-sensitive Lick indices plotted against log(σ) for iron and calcium sensitive features. <Fe> is the average of Fe5270 and Fe5335. Trend-lines and data-point symbols are here as described in Figure 1a.

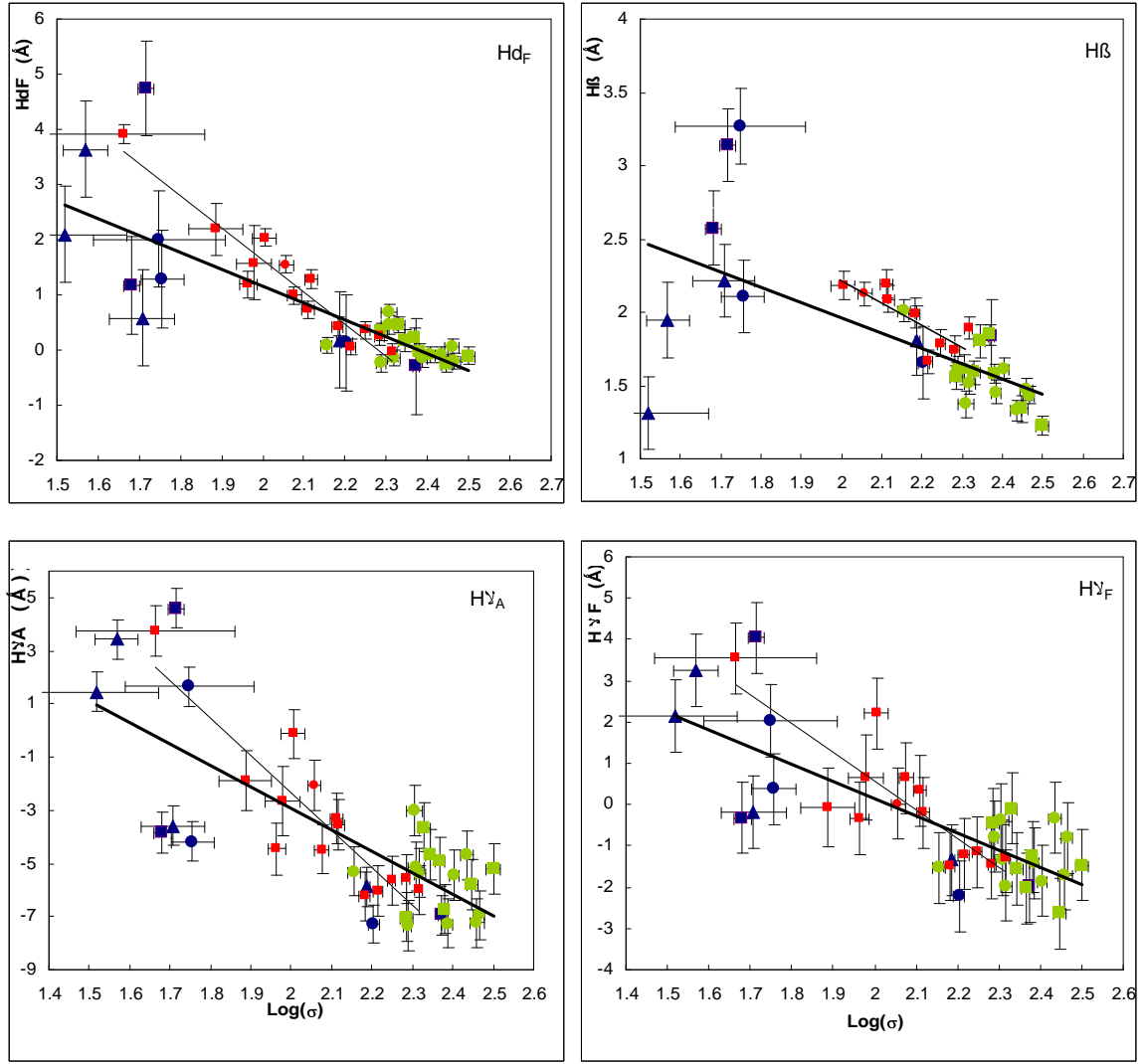

**Figure 2.** Predominantly age sensitive indices versus log(σ). Trend-lines and data-point symbols are here as described in Figure 1a.

| Index | Slope | Intercept | r | r(NTT only) | r (WHT only) |
|---|---|---|---|---|---|
| Ca4227 | 1.04 | -1.16 | 0.75 | 0.68 | 0.21 |
| Fe5015 | 3.71 | -2.81 | 0.76 | 0.64 | 0.02 |
| <Fe> | 1.92 | -1.53 | 0.8 | 0.71 | -0.19 |
| Fe5406 | 1.21 | -0.8 | 0.86 | 0.79 | -0.21 |
| | | | | | |
| <CN> | 0.24 | -0.42 | 0.88 | 0.93 | 0.21 |
| $C_2$4668 | 8.23 | -11.4 | 0.9 | 0.88 | -0.12 |
| Mg1 | 0.16 | -0.23 | 0.93 | 0.94 | 0.35 |
| Mg2 | 0.31 | -0.41 | 0.94 | 0.90 | 0.51 |
| Mgb | 4.66 | -5.93 | 0.94 | 0.88 | 0.60 |
| | | | | | |
| H$\delta$F | -3.06 | 7.28 | -0.79 | -0.68 | -0.43 |
| H$\gamma$A | -8.08 | 13.24 | -0.77 | -0.76 | -0.19 |
| H$\gamma$F | -4.17 | 8.47 | -0.79 | -0.79 | -0.21 |
| Hß | -1.04 | 4.05 | -0.64 | -0.26 | -0.71 |

**Table 4.** Correlations between Lick indices and log($\sigma$), for the current NTT data (mainly LLEs) and WHT early-type galaxies (from PS02). Number of galaxies used to form these values is 10 from the current NTT data set, 17 from the WHT data set (PS02). Table columns are described in main text.

In Table 4 columns are firstly index name, gradient and intercept for combined data sets (current NTT data plus early-type galaxies from PS02). Next is the Pearson's r for the early-type galaxies in PS02 plus current NTT data combined. Pearson's r for the NTT data alone, then WHT early-type galaxies alone (from PS02) follow, in the last two columns of Table 4 respectively. All the correlations for the combined data are statistically significant, with r>0.73, except for Hß due to its large scatter in the NTT data. The correlations for the current NTT data alone are similar, though slightly less significant in some cases. None of the correlations for the early-type galaxies in PS02 alone are significant owing to the narrow range in log($\sigma$) for those data.

In Figure 1a, Figure 1b and Figure 2 the lines of best fit for late and early-type galaxies are quite close to one another. The question of whether the LLEs compare well with the late-type galaxies from PS02 was quantitively investigated from the r coefficients of the correlations of $\sigma$ versus various Lick indices using the NTT data combined with the PS02 late-type galaxies. The correlations found are given in Table 5. The last column of Table 5 is for PS02 data for late-type galaxies only. All correlations in Table 5 are statistically significant (r>0.73), again except for the case of Hß.

| Index | Slope | Intercept | r | r (WHT only) |
|-------|-------|-----------|------|------|
| Ca4227 | 1.36 | -1.78 | 0.77 | 0.90 |
| Fe5015 | 4.97 | -4.94 | 0.78 | 0.95 |
| <Fe> | 2.48 | -2.53 | 0.82 | 0.96 |
| Fe5406 | 1.51 | -1.42 | 0.86 | 0.97 |
| | | | | |
| <CN> | 0.25 | -0.47 | 0.85 | 0.91 |
| $C_2$4668 | 9.89 | -14.72 | 0.91 | 0.96 |
| Mg1 | 0.18 | -0.26 | 0.92 | 0.94 |
| Mg2 | 0.34 | -0.46 | 0.93 | 0.97 |
| Mgb | 4.95 | -6.53 | 0.93 | 0.98 |
| | | | | |
| HdF | -3.83 | 8.93 | -0.75 | -0.95 |
| HγA | -10.96 | 18.84 | -0.79 | -0.88 |
| HγF | -5.40 | 10.95 | -0.78 | -0.84 |
| Hß | -0.67 | 3.42 | -0.38 | -0.81 |

**Table 5.** Correlations between Lick indices and $\log(\sigma)$, for the NTT data set and the 15 late-type WHT galaxies used in PS02 for correlations. Columns as described in text.

Therefore the LLEs show about the same level of correlation with the early-type galaxies as with the late-type galaxies from PS02. There is no preference. Figure 1 (a & b) shows that the LLEs lie closer to the spiral bulges than they do to the extrapolation of trends for more luminous early-type galaxies (from K00, for $75 < \sigma < 375$ km s$^{-1}$). Therefore, the slope of the trend for early-type galaxies is seen to steepen at lower velocity dispersions. A hint of this was given by CRC03 (their figure 5) whose data showed a large downward scatter at low velocity dispersions. Moderately large scatter is also found in the present sample of LLEs, in excess of the errors.

Table 3: Lick Index Measurements of the NTT data set. Errors are given on the 2$^{nd}$ row corresponding to each galaxy.

| Galaxy | Hδ_F (Å) | CN_1 (mag) | CN_2 (mag) | Ca4227 (Å) | G4300 (Å) | Hγ_A (Å) | Hγ_F (Å) | Fe4383 (Å) | Ca4455 (Å) | Fe4531 (Å) | C_4668 (Å) | Hβ (Å) | Fe5015 (Å) | Mg_1 (mag) | Mg_2 (mag) | Mg_b (Å) | Fe5270 (Å) | Fe5335 (Å) | Fe5406 (Å) |
|---|---|---|---|---|---|---|---|---|---|---|---|---|---|---|---|---|---|---|---|
| ESO 118-34 | 4.743 | -0.1195 | -0.077 | 0.082 | -0.153 | 4.592 | 4.040 | 1.347 | -0.084 | 1.807 | -0.284 | 3.139 | 1.527 | 0.0227 | 0.0639 | 1.139 | 0.912 | 0.799 | 0.674 |
|  | 0.869 | 0.0827 | 0.0450 | 0.158 | 0.186 | 0.740 | 0.871 | 3.335 | 0.196 | 2.471 | 0.201 | 0.252 | 0.419 | 0.0021 | 0.0024 | 0.097 | 0.094 | 0.095 | 0.512 |
| ESO 157-30 | 2.007 | -0.0500 | -0.0098 | 0.298 | 1.607 | 1.657 | 2.007 | 0.315 | 0.255 | 2.073 | 1.504 | 3.270 | 2.337 | 0.0194 | 0.0679 | 1.009 | 1.245 | 0.692 | 1.158 |
|  | 0.877 | 0.0450 | 0.0453 | 0.174 | 0.186 | 0.749 | 0.874 | 3.339 | 0.196 | 2.473 | 0.247 | 0.257 | 0.426 | 0.0024 | 0.0027 | 0.106 | 0.106 | 0.108 | 0.513 |
| ESO 358-59 | 0.574 | 0.0828 | 0.0198 | 0.874 | 4.251 | -3.592 | -0.192 | 3.155 | 0.431 | 4.275 | 3.726 | 2.213 | 4.502 | 0.0575 | 0.1632 | 2.964 | 2.551 | 2.515 | 0.510 |
|  | 0.872 | 0.0827 | 0.0451 | 0.163 | 0.186 | 0.744 | 0.871 | 3.335 | 0.196 | 2.471 | 0.186 | 0.246 | 0.405 | 0.0020 | 0.0023 | 0.089 | 0.081 | 0.076 | 0.511 |
| NGC1331 | 1.289 | 0.0001 | 0.0513 | 1.327 | 5.323 | -4.175 | 0.370 | 5.274 | 0.903 | 3.648 | 5.338 | 2.109 | 5.849 | 0.0749 | 0.1736 | 2.710 | 2.792 | 2.251 | 2.224 |
|  | 0.877 | 0.0828 | 0.0453 | 0.171 | 0.186 | 0.749 | 0.873 | 3.336 | 0.196 | 2.472 | 0.205 | 0.249 | 0.411 | 0.0022 | 0.0025 | 0.096 | 0.090 | 0.089 | 0.508 |
| NGC 1374 | 0.137 | 0.1430 | 0.1875 | 1.158 | 6.351 | -7.281 | -2.223 | 5.449 | 1.475 | 3.982 | 7.471 | 1.660 | 5.786 | 0.1699 | 0.3281 | 5.069 | 2.959 | 2.684 | 1.682 |
|  | 0.870 | 0.0826 | 0.0449 | 0.153 | 0.186 | 0.739 | 0.871 | 3.332 | 0.196 | 2.469 | 0.126 | 0.246 | 0.393 | 0.0016 | 0.0018 | 0.071 | 0.054 | 0.037 | 0.510 |
| NGC 1375 | 1.174 | -0.0307 | 0.0023 | 1.218 | 4.632 | -3.836 | -0.319 | 4.428 | 1.131 | 4.175 | 4.378 | 2.574 | 5.712 | 0.0607 | 0.1887 | 2.799 | 2.707 | 2.570 | 1.675 |
|  | 0.879 | 0.0828 | 0.0453 | 0.170 | 0.186 | 0.750 | 0.874 | 3.336 | 0.196 | 2.471 | 0.200 | 0.251 | 0.408 | 0.0021 | 0.0024 | 0.093 | 0.093 | 0.082 | 0.508 |
| NGC 1411 | 0.180 | 0.0988 | 0.138 | 1.066 | 5.479 | -5.875 | -1.340 | 5.494 | 1.236 | 3.786 | 7.141 | 1.806 | 5.273 | 0.1272 | 0.2579 | 4.212 | 3.306 | 3.093 | 0.920 |
|  | 0.868 | 0.0826 | 0.0449 | 0.152 | 0.186 | 0.737 | 0.869 | 3.332 | 0.196 | 2.469 | 0.122 | 0.240 | 0.388 | 0.0016 | 0.0018 | 0.070 | 0.060 | 0.035 | 0.510 |
| NGC 2328 | 2.093 | -0.0052 | -0.0398 | 0.148 | 0.888 | 1.456 | 2.152 | 0.984 | -0.056 | 1.728 | 0.193 | 1.310 | 2.553 | 0.0308 | 0.0759 | 1.407 | 1.342 | 1.174 | 2.069 |
|  | 0.869 | 0.0827 | 0.0449 | 0.156 | 0.186 | 0.739 | 0.870 | 3.334 | 0.196 | 2.470 | 0.172 | 0.246 | 0.402 | 0.0019 | 0.0022 | 0.086 | 0.077 | 0.072 | 0.508 |
| NGC 2784 | -0.290 | 0.1436 | 0.1920 | 1.462 | 5.690 | -6.954 | -1.983 | 6.362 | 1.507 | 4.207 | 8.846 | 1.834 | 5.870 | 0.1749 | 0.3419 | 4.998 |  | 2.707 |  |
|  | 0.871 | 0.0827 | 0.0450 | 0.156 | 0.186 | 0.742 | 0.872 | 3.333 | 0.196 | 2.470 | 0.143 | 0.247 | 0.396 | 0.0017 | 0.0020 | 0.075 | 0.053 | 0.047 |  |
| NGC 59 | 3.643 | -0.0737 | -0.0487 | 0.108 | -0.308 | 3.443 | 3.258 | 1.250 | -0.36 | 1.328 | 0.156 | 1.946 | 0.974 | 0.0258 | 0.0320 | 0.510 | 0.692 | 0.703 | 0.888 |
|  | 0.873 | 0.0827 | 0.0451 | 0.166 | 0.186 | 0.745 | 0.873 | 3.336 | 0.196 | 2.472 | 0.221 | 0.259 | 0.449 | 0.0023 | 0.0025 | 0.101 | 0.100 | 0.102 | 0.512 |

# 6 DERIVED GALAXY AGES AND ABUNDANCES

Ages and abundances of galaxies in the NTT data set were estimated using SSP fitting software written by Proctor (2002), modified here to handle the NTT data set. This SSP fitting software searches 85 discrete steps of Log(Age) and 111 discrete steps of [Fe/H] to select SSPs from the arrays of Vazdekis 1999 (see PS02) with adjustments from Tripicco and Bell (1995) to allow for abundance ratio enhancements using the parameter [$\alpha$/Fe]. The software searches through 37 possible enhancements (to allow for $\alpha$-elements to Fe-peak elements ratio different from solar). This produces a search of 349095 possible combinations of age (from 1.5 to 16.8 Gyr), [Fe/H] (0.01 to 5.6 times solar) and enhancement (about 0.5 to 4 times solar abundance ratio). Chi-squared fits of models to typically 19 Lick indices were used to determine the best fitting age, [Fe/H] and [$\alpha$/Fe] SSP parameters and their errors.

The measured LLE line strengths (Table 3) were used as inputs to this SSP fitting program, using the "Fe-" method in PS02, to derive luminosity weighted Log(Age), [Fe/H] and [$\alpha$/Fe] parameters. This method elevates all elements ($\alpha$-capture plus Fe-peak) using Tripicco & Bell (1995) models, then drops the Fe-peak elements down to generate non-solar [$\alpha$/Fe] ratios in the model SSPs. For errors on the derived parameters, contours were used in chi-squared space at the 1-sigma level above the best fit, allowing for 3 interesting parameters. This appeared to give reasonable error estimates for all the galaxies.

For ESO118-34, NGC2328 and NGC59 the best fits including Hß were substantially poorer, therefore Hß was excluded for the SSP fits to those galaxies, since it is most strongly affected by uncertainties in emission line filling. Best fits with and without Hß for the other NTT galaxies gave robust results, within the errors on the derived parameters. As an additional test, the 5 lines most affected by emission line filling were excluded from the fits. The best fit SSPs did not differ significantly from the values found including these lines, however the errors slightly increased. These tests with and without emission line affected features indicate that we can be confident in the derived ages and abundances. Determinations of [$\alpha$/Fe] were found to be particularly robust in these tests.

As an illustration of the differences found between giant early-type galaxies and some of the low luminosity early-type galaxy properties revealed in this current work, two spectra are compared in Figure 3. These are for the control giant lenticular galaxy NGC 2784 (which is metal-rich, $\alpha$-enhanced and old) and the low luminosity elliptical ESO157-30 (which is metal-poor, $\alpha$-deficient and has a younger stellar population), both from the current sample of NTT data. Figure 3 shows a qualitative illustration of the evidence for these differences in derived parameters directly from the relative strengths of the lines in the spectra themselves. Relative to NGC 2784 the hydrogen lines (e.g. Hß at 4861Å) are much stronger in ESO157-30 indicative of its younger age population, whilst the metal sensitive lines (e.g. Mgb at 5183Å, Fe sensitive lines at 5270 and 5335Å) are much weaker, and particularly so for the magnesium sensitive lines. This illustrates directly the low metallicity and low magnesium-to-iron ratio found in ESO157-30 compared to the giant elliptical NGC 2784.

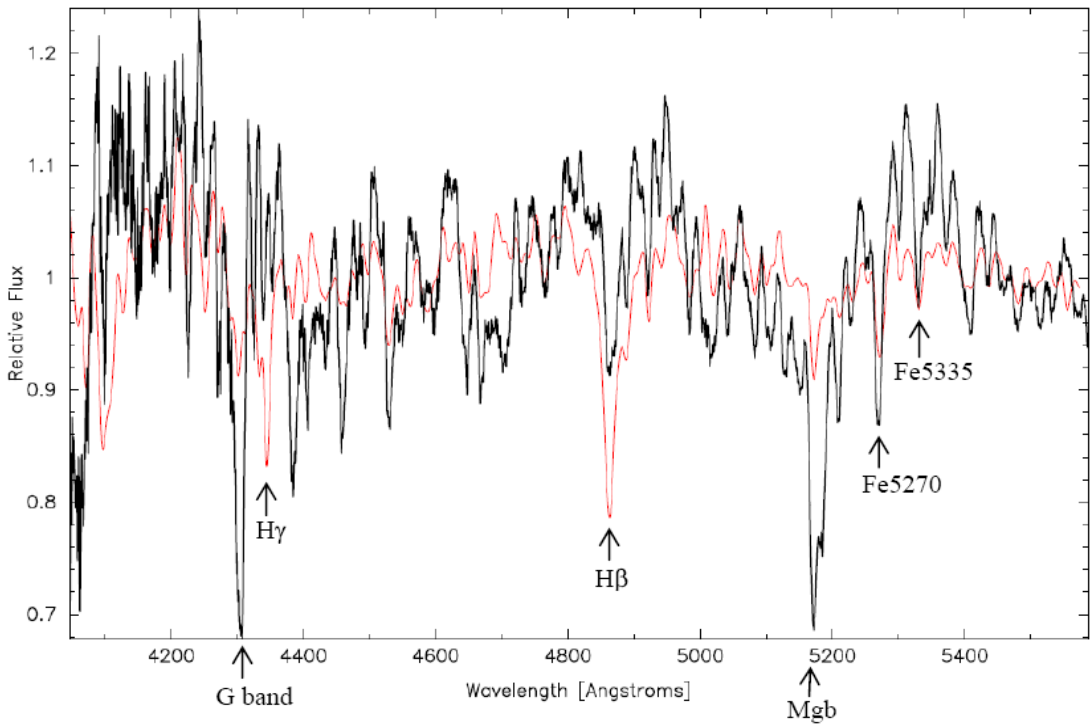

**Figure 3.** Comparison of spectra for a luminous lenticular galaxy NGC 2784 (in black) with a low luminosity elliptical ESO157-30 (in red) illustrating the differences in line strength for age and metallicity sensitive features from which we derive these mean population parameters. These NTT spectra were divided by a fifth order polynomial to remove broad continuum trends, shifted to rest wavelengths and the ESO157-30 spectrum was Gaussian smoothed to match the velocity dispersion of NGC 2784 for direct comparison of spectral feature strengths. Some key indices are labelled.

The derived luminosity weighted mean ages, [Fe/H] and [α/Fe] values for these galaxies are given in Table 6. These LLEs have relatively young ages (<3Gyrs); a wide range of metallicities (-1.3<[Fe/H]<+0.1 dex) and low α-element to iron peak abundance ratios (-0.3<[α/Fe]<+0.075). This is similar to what was found by van Zee, Barton & Skillman (2004), at lower signal-to-noise, for an ensemble of 16 dwarf ellipticals in the Virgo Cluster. Their data indicate low metallicities ([Fe/H]<-0.33) and suggest sub-solar abundance ratios. They used 4 indices and illustrated their results in index-index plots. No attempt at simultaneous fitting or interpretation between index plots was made in that paper. Our data improves substantially on this by deriving parameters for individual galaxies, using $\chi^2$ fitting to many line-strength indices (Table 6).

For the current sample of LLEs plus data from PS02, plots were made of log(Age), [Fe/H] and [α/Fe] against log(σ) and regression line fits to these plots are given in Table 7. In Figure 4, galaxy ages are plotted versus σ, for the NTT and WHT data sets. Ages and abundances for the WHT galaxies were taken from Table 10 of PS02. In Figure 5, [Fe/H] is plotted against log(σ). In Figure 6, [α/Fe] is plotted against log(σ). Figure 7 shows [Fe/H] plotted against log(Age) and Figure 8 shows [α/Fe] plotted against log(Age). These figures illustrate the relative properties of the LLEs.

| Galaxy | Log(σ) | Log(σ) Error | Log(Age) | Log(Age) error | [Fe/H] | [Fe/H] error | [α/Fe] | [α/Fe] error |
|---|---|---|---|---|---|---|---|---|
| NGC59 | 1.568 | 0.054 | 0.175 | 0.031 | -1.350 | 0.075 | -0.300 | 0.075 |
| ESO358-59 | 1.708 | 0.078 | 0.463 | 0.088 | -0.025 | 0.088 | -0.175 | 0.050 |
| NGC1375 | 1.681 | 0.021 | 0.463 | 0.081 | 0.100 | 0.063 | -0.150 | 0.038 |
| NGC1331 | 1.756 | 0.053 | 0.400 | 0.056 | 0.100 | 0.050 | -0.075 | 0.050 |
| ESO157-30 | 1.748 | 0.161 | 0.475 | 0.119 | -1.175 | 0.113 | -0.200 | 0.113 |
| ESO118-34 | 1.716 | 0.019 | 0.175 | 0.019 | -1.175 | 0.088 | 0.075 | 0.138 |
| NGC2328 | 1.519 | 0.152 | 0.338 | 0.106 | -0.975 | 0.163 | -0.150 | 0.100 |
| NGC1411 | 2.121 | 0.006 | 0.425 | 0.063 | 0.425 | 0.038 | 0.075 | 0.013 |
| NGC1374 | 2.204 | 0.013 | 1.013 | 0.044 | 0.050 | 0.025 | 0.300 | 0.013 |
| NGC2784 | 2.373 | 0.008 | 0.863 | 0.075 | 0.225 | 0.025 | 0.275 | 0.013 |

**Table 6.** NTT data set ages, metallicities [Fe/H] and alpha-to-iron abundance ratios [α/Fe]. Units are kms$^{-1}$ for σ, Gyr for Age and dex for abundances.

| Y-axis | X-axis | Data | No. of points | Slope | Intercept | R |
|---|---|---|---|---|---|---|
| Log(Age) | Log(σ) | All | 41 | 0.75 | -0.99 | 0.79 |
| [Fe/H] | Log(σ) | All | 41 | 1.14 | -2.43 | 0.70 |
| [α/Fe] | Log(σ) | All | 41 | 0.47 | -0.85 | 0.81 |
| [Fe/H] | Log(Age) | All | 41 | 0.68 | -0.41 | 0.40 |
| *[Fe/H]* | *Log(Age)* | *WHT E/S0* | *17* | *-0.60* | *0.68* | *-0.81* |
| [α/Fe] | Log(Age) | All | 41 | 0.41 | -0.10 | 0.68 |
| *[α/Fe]* | *Log(Age)* | *Not NTT* | *31* | *0.23* | *0.06* | *0.71* |

**Table 7.** Correlations between Log(σ) and Log(Age), [Fe/H] and [α/Fe], for NTT plus WHT data. The correlation between [α/Fe] and Log(σ) is very significant including all these data for E, S bulges and LLE galaxies. The correlation between [Fe/H] and Log(Age) is only significant for the bright E/S0 galaxies from the WHT data (from PS02, with restricted σ range). The correlation between [α/Fe] and Log(Age) is quite significant for all the WHT data from PS02, however this is slightly weakened when the LLEs are included, since they are offset from the main trend (Figure 8). Units are Age (Gyr), σ (km s$^{-1}$), and dex for abundance ratios [Fe/H] and [α/Fe].

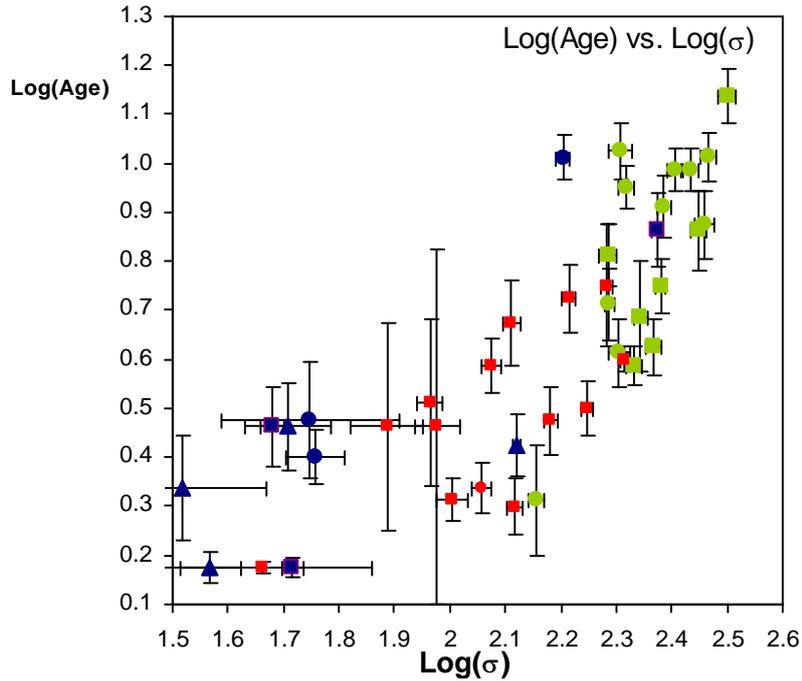

**Figure 4.** Log(Age) versus Log(σ), for NTT and WHT data sets. The units of σ are kms[-1]; those of Age are Gyr. Colours of data points are as in Figure 1a. To summarise: dark blue points are from the NTT data set (mainly LLEs), red points are late-type galaxies from the WHT data set and light green points are bright, early-type (E and S0) galaxies from the WHT data set.

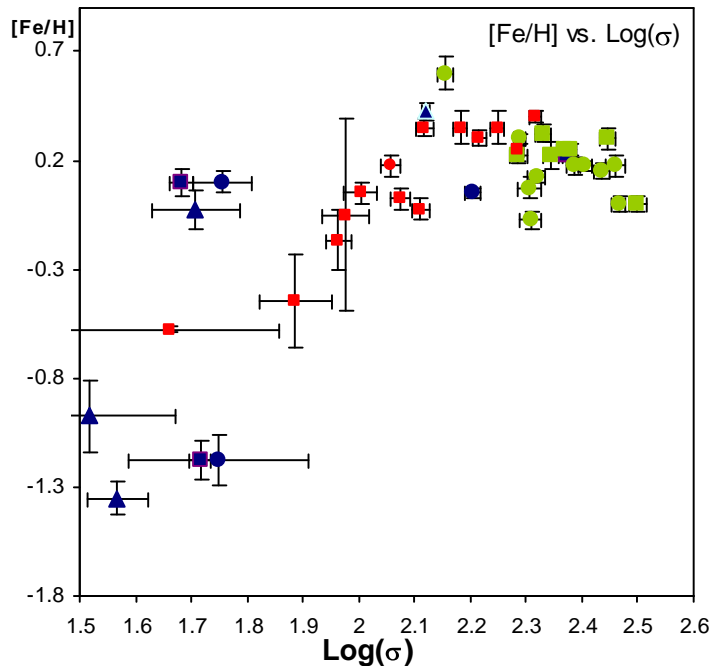

**Figure 5.** [Fe/H] versus Log(σ), for NTT and WHT data sets. This figure highlights the very low metallicities of some LLEs. The units of σ are kms[-1]. Colours of data points are as described in Figure 4.

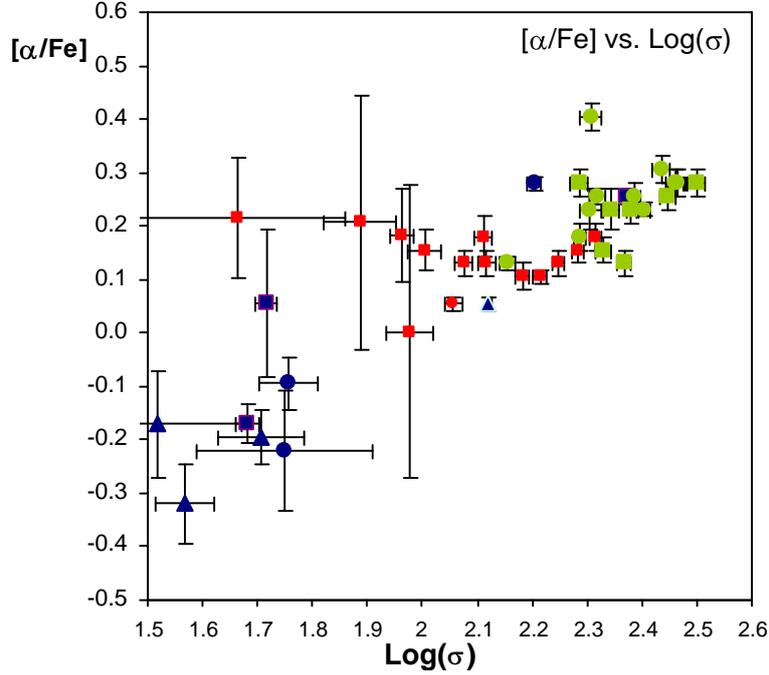

**Figure 6.** [α/Fe] versus Log(σ), for NTT and WHT data sets. The LLEs appear to continue the trend found for more luminous E and S galaxies down to lower Log(σ) and [α/Fe] values. Brighter ellipticals used as controls in the NTT data set are also consistent with this trend. The units of σ are kms$^{-1}$. Colours of data points are as described in Figure 4.

These results show that the luminosity weighted mean ages of LLEs are comparable to those of spiral bulges with low velocity dispersions (Figure 4). The control cases (NGC 2784 and NGC 1374) lie in expected locations for their velocity dispersions, compared with the previous results. LLEs have much younger mean ages than bright ellipticals. This is consistent with the findings of others: e.g. for dwarf galaxies in Coma (Poggianti et al. 2001); evidence of extended star formation, from resolved star data for dwarf early-type galaxies in the Local Group (Dolphin et al., 2005) and for low luminosity dE and dSph galaxies (-11.3>M$_V$>-14.2) in M81 (Da Costa et al. 2007). Some of the LLEs shown here have very low metallicities ([Fe/H] <-1.0), whereas others, of similar velocity dispersion, have comparable metallicities to those in spiral bulges ([Fe/H]~0.0). This is highlighted in Figure 5 and in Figure 7, which shows the clear offset of the metal-poor LLEs from the log(Age) versus [Fe/H] anticorrelation found in bright ellipticals (PS02). The metal-poor group (typically [Fe/H]<-1.0) includes both lenticular and elliptical Hubble types (from Table 1) and are in lower density environments rather than in the Fornax cluster. However, larger samples are needed to accurately probe environmental trends. This bimodality in metallicity (Figure 7) was suggested in the work of Poggianti et al. (2001), where they found ages and metallicities from lower S/N spectra of a large number of non-emission line, dwarf galaxies (M$_B$ > -17.3) in the Coma cluster and suggested that higher S/N were needed to confirm their finding. They did not look at abundance ratios. Here we find support for their suggested bimodality in young low luminosity, early-type galaxies. For a large sample of galaxies Gallazzi et al (2005) showed that low mass early-type galaxies have a large range of metallicities with a suggestion of bimodality (see their figure 12, top left plot), from probability density mapping. Recently Michielsen et al. (2007) derived population parameters for 24 dwarf E/S0 galaxies in the Virgo cluster, using 4 Lick indices. Their results also indicate a large range in metallicity, going down to ~1/10[th] of solar. They find a broad age range and a scatter of abundance ratios around solar (their figures 7 and 8). Their spectral resolution is poorer and their derived parameter error bars are larger than for the current sample, however they do cover a similar luminosity range.

To clarify the reality of the significantly different [Fe/H] measurements from the current NTT data, Figure 9 compares the spectra of two LLEs: ESO118-34 (with [Fe/H] = -1.175) and ESO358-59 (with [Fe/H] = -0.025). This comparison focuses on a region of the spectrum containing magnesium and iron sensitive metal absorption lines. Figure 9 clearly illustrates the low [Fe/H] in ESO118-34 when compared to ESO358-59.

Abundance ratios in LLEs show a continuation of the trend seen for brighter galaxies, with [$\alpha$/Fe] decreasing steadily to sub-solar values for the LLEs (Figure 6). This is in contrast to globular clusters and giant ellipticals in different environments, which all have enhanced [$\alpha$/Fe] ratios (e.g. Mendel, Proctor & Forbes 2007; Mendes de Oliveira et al. 2005; K00). Going to lower luminosity systems, dSph galaxies show a wide range of [$\alpha$/Fe] values, from super-solar to sub-solar (Carigi, Hernandez & Gilmore 2002; Venn et al., 2004), including some with [Fe/H] and [$\alpha$/Fe] values similar to those found in the low metallicity LLEs of the current work. Venn et al. (2004) found that none of the stellar populations in our Galaxy could be formed from dSphs since [$\alpha$/Fe] ratios in those galaxies are too low, indicative of slow chemical evolution. They suggested that higher mass dwarf galaxies might have had more rapid chemical evolution. However, the results for LLEs presented here show that they too have [$\alpha$/Fe] values that are low compared with other stellar populations. It is clear from Figure 6 that LLEs as a class have low [$\alpha$/Fe] values.

Overall there are significant trends for Age and [$\alpha$/Fe] to increase with velocity dispersion, which may be due to an underlying trend between Age and [$\alpha$/Fe]. However, Figure 8 shows that, again the LLEs are offset from the main trend, with the control galaxies lying in their expected locations, hidden behind points from previous data.

Therefore the LLEs have systematically younger mean ages than brighter galaxies, they have systematically lower [$\alpha$/Fe] ratios and they fall into two categories of metallicities, one at [Fe/H]<-1.0 and one around [Fe/H]~0.0. These measurements constrain the types of star formation histories that LLEs could have undergone, even though we are only currently measuring luminosity weighted average properties. For example, an early burst of star formation, cut off by feedback, would not produce these low [$\alpha$/Fe] values, or young ages. The bimodal behaviour in metallicity might suggest two different types of star formation histories, and hence formation processes, are at work in LLEs. This warrants further study in future to use these results to investigate possible histories of formation.

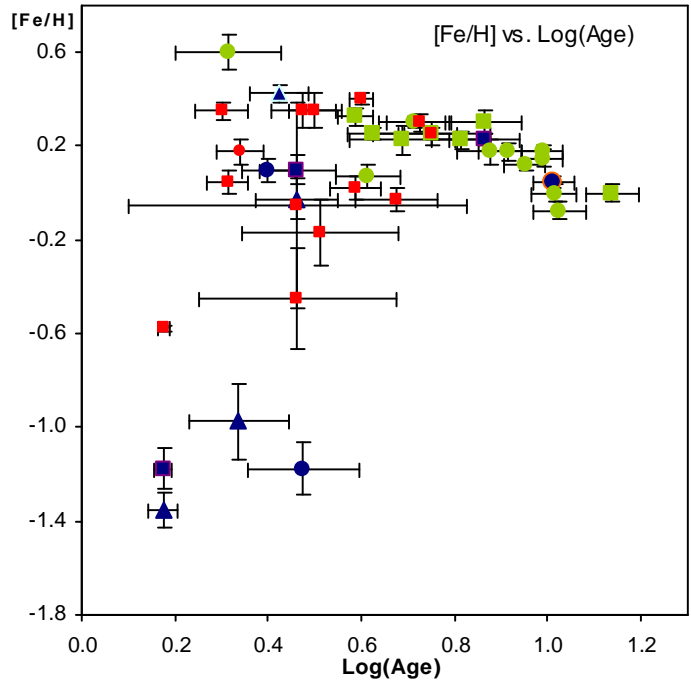

**Figure 7.** [Fe/H] is plotted against Log(Age). The anti-correlation between age and metallicity seen in luminous ellipticals (green points) is not followed by the LLEs (dark blue points), which can show dramatically different metallicities. Units of Age are Gyr. Colours of data points are as described in Figure 4.

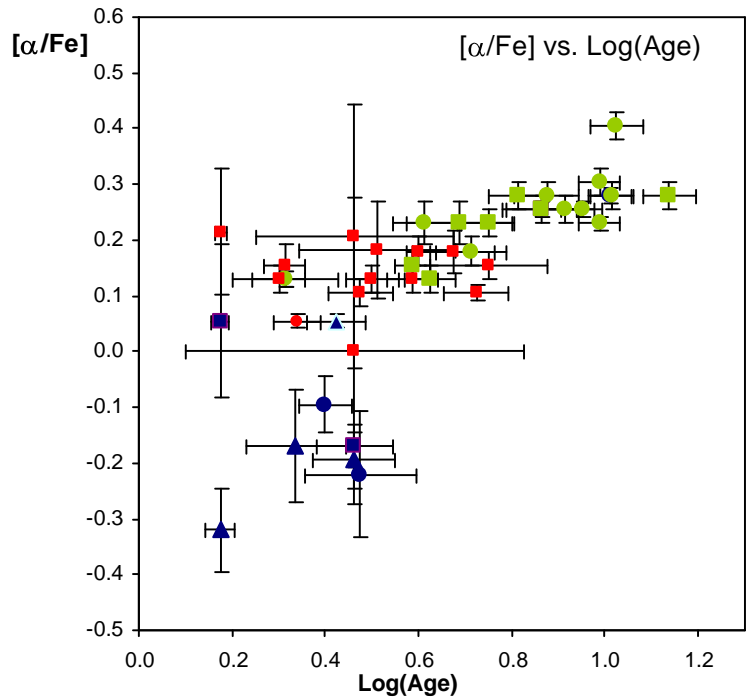

**Figure 8.** [α/Fe] is plotted against Log(Age). This figure shows that some of the LLEs do not follow the trend found for more luminous galaxies. Units of Age are Gyr. Colours of data points are as described in Figure 4.

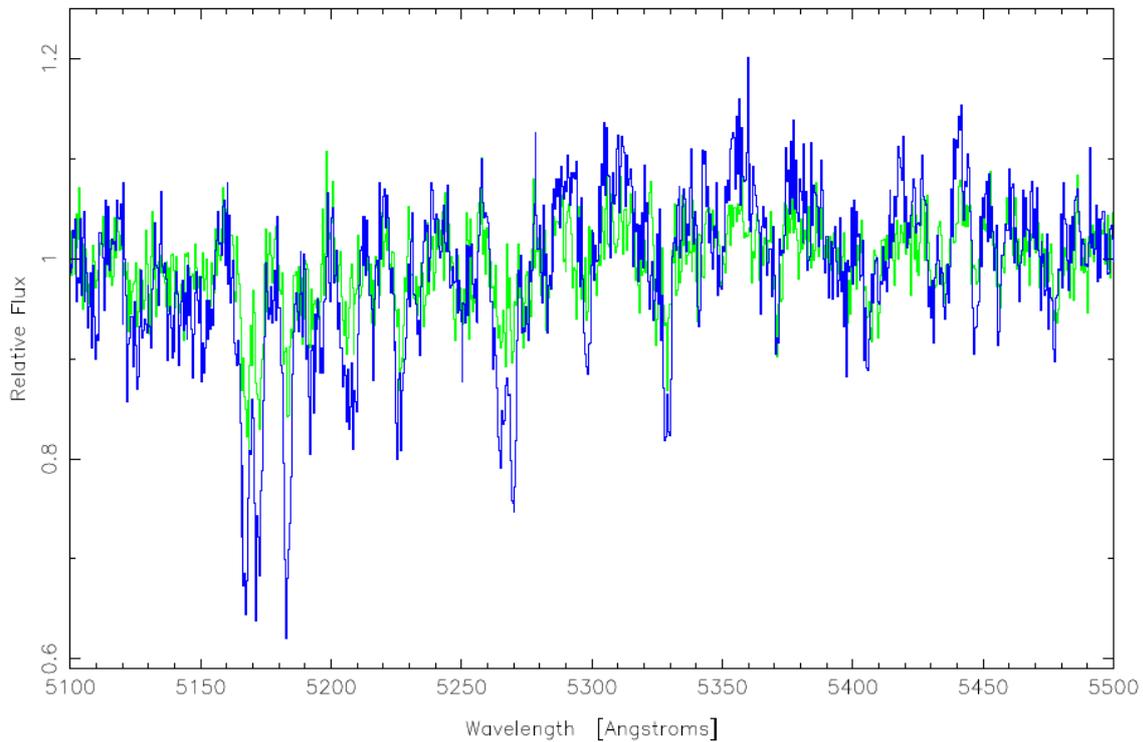

**Figure 9.** Comparison of NTT spectra around Mgb, Fe5270 and Fe5335 region for two LLEs with the same σ, within errors, but very different abundances: ESO118-34 (with <1/10th solar [Fe/H] and ~ solar [α/Fe]) and ESO358-59 (with ~solar [Fe/H] and sub-solar [α/Fe]). These spectra are shifted to rest wavelengths, and divided by a third order polynomial to remove broad continuum trends. The stronger lines in ESO358-59 (dark blue) can clearly be seen in comparison to the weaker lines in ESO118-34 (light green).

## 7. CONCLUSIONS

A sample of 10 early-type galaxies were observed with the NTT EMMI spectrograph, 8 of which are low luminosity ($M_B$ fainter than 18th mag.), in order to study the stellar populations in early-type galaxies at low velocity dispersions. Two other galaxies observed were not analysed here. The measured velocity dispersions range from 33 to 154 km s-1 for the LLEs, covering a range that has not previously been well explored at high spectral signal-to-noise. These data were compared to published kinematics and line-strength indices for ellipticals, lenticulars and spiral bulges from PS02.

The results show that the LLEs have similar line strengths to spiral bulges of similar velocity dispersion. LLEs lie closer to the spiral bulges than they do to extrapolation of the trends for more luminous early-type galaxies (from K00). Therefore, the slope of the trend for early-type galaxies is seen to steepen at lower velocity dispersions. These results are reflected in the derived ages and metallicities of galaxies at different velocity dispersions and hence will provide constraints on cosmological models of galaxy formation. The previously determined relations for early-type galaxies, based mainly on brighter ellipticals, do not extend well to the properties of LLEs, in line-strength versus velocity dispersion planes.

The LLEs show younger ages than massive elliptical and lenticular galaxies. They have similar ages to those estimated for spiral bulges from Lick indices. The ages of these LLEs range from <1.5 to 3 Gyr. They show a wide metallicity distribution ([Fe/H] from -1.35 to +0.4, which may be bimodal. The LLEs have sub-solar to ~solar [$\alpha$/Fe] ratios, ranging from (-0.3 to +0.075).

These data reveal that LLEs have very different abundance distributions compared to other unresolved stellar populations studied so far in brighter galaxies, and to resolved populations in our own Galaxy. The LLEs with the lowest [$\alpha$/Fe] ratios tend to be those with the lowest metallicities. These abundance patterns, with low metallicity and low [$\alpha$/Fe] ratio are different to those found in most other stellar population, for example the local solar neighbourhood, globular clusters, galaxy bulges or bright elliptical and lenticular galaxies. They are most similar to abundance patterns found in some dwarf spheroidal galaxies. Thus we have extend the conclusions regarding dSphs, to show that these low luminosity galaxies, more luminous than dSphs, also cannot have formed the building blocks for stellar populations in luminous early-type galaxies through hierarchical mergers. The abundances patterns in both dSphs and LLEs are too different from giant ellipticals for this to have occurred. These data thus provide a challenge to model and will constrain the possible star formation histories that could have occurred in LLEs, as well as constraining the types of progenitors that led to more luminous galaxies. Possible histories for these LLEs will be explored with chemical evolution models in a future paper.

## ACKNOWLEDGEMENTS

This work is based on observations obtained with the New Technology Telescope run by ESO at the La Silla Observatory under programme ID 072.B-0641. MSN acknowledges the support of a PPARC studentship grant. Thanks to Helen Cammack of Runshaw College, Leyland for help in producing Figure 3 and for helpful comments from Leticia Carigi and an anonymous referee. The authors acknowledge the usage of the HyperLeda database (http://leda.univ-lyon1.fr). This research has also made use of the SIMBAD database, operated at CDS, Strasbourg, France (http://simbad.u-strasbg.fr).

# APPENDIX A: CALIBRATIONS

In this Appendix some details of calibration to the Lick scale are given. Table A1 gives the flux and Lick standard stars observed for flux, kinematic and index calibrations.

| Stars | Spectral Type | V Mag | Exp (secs) | Air Mass |
|---|---|---|---|---|
| **Flux Calibrators:** | | | | |
| HR8634 | B8V | 3.40 | 10 | 1.33 |
| HD49798 | sdO6p | 8.27 | 10 to 50 | 1.10 |
| HR3454 | B3V | 4.27 | 10 to 30 | 1.20 |
| **Lick Calibrators:** | | | | |
| HR3845 | K2.5III | 3.90 | 2 to 5 | 1.23 |
| HR8841 | K0III | 4.21 | 10 | 1.07 |
| HR72 | G0V | 6.40 | 10 to 30 | 1.10 |
| HR203 | G2V | 6.15 | 30 | 30 |
| HR3994 | K0III | 3.61 | 5 | 1.18 |
| HR296 | K0III | 5.40 | 10 | 1.19 |
| HR8924 | K3III | 6.30 | 10 | 1.11 |
| HD211038 | K0/K1V | 6.54 | 10 | 1.09 |
| HR509 | G8V | 3.50 | 5 | 1.18 |
| HD219617 | F8IV | 8.16 | 50 to 100 | 1.05 |
| HR2574 | K4III | 4.08 | 5 | 1.14 |
| HR2701 | K0III | 4.92 | 10 | 1.2 |
| HR2970 | G9III | 3.93 | 10 | 1.09 |
| HR695 | G2V | 5.19 | 10 | 1.24 |
| HR1136 | K0IV | 3.51 | 5 | 1.26 |
| HD064606 | G8V | 7.44 | 100 | 1.14 |
| HR4287 | K1III | 4.07 | 10 | 1.23 |

**Table A1.** Calibration stars observed with NTT. Spectral type, V magnitude, exposure time and air mass are given.

To correct the NTT observations to the spectral resolution of the LICK/IDS system, the line-strength software (BANSI) was adjusted so as to handle NTT data with dispersion of 0.41 Å per pixel. The Lick spectral resolution, $\sigma_L$, is given in Table A2 (Worthey & Ottaviani 1997) and the instrumental broadening, $\sigma_I \sim 1.83$ Å was measured from the width of arc lines. Those galaxies whose spectra are broadened by less than $\sigma_L$ (all except for the control case NGC 2784) had their spectra broadened up to the Lick resolution. This was done by convolution with a Gaussian of width $\sigma_B$ given by equation A1:

$$\sigma_B = (\sigma_L{}^2 - \sigma^2 - \sigma_I{}^2)^{1/2}. \tag{A1}$$

Where $\sigma$ is the measured velocity dispersion in each galaxy.

   To measure the effect of velocity dispersion broadening, where the total broadening is greater than $\sigma_L$ (i.e. for NGC 2784 only in our sample) the various Lick star spectra were smoothed to varying widths in the range of $\sigma_c = 0$ to 300 kms$^{-1}$, in intervals of 20 kms$^{-1}$. From these smoothed spectra, correction factors $C_i(\sigma_c)$ could be determined, following the prescription laid down in Appendix A of PS02. For molecular indices (i.e. those with units of magnitudes) and line strengths of H$\gamma$ and H$\delta$ (which can straddle zero) correction factors were determined from Equation A2:

$$C_i(\sigma) = I_{orig} - I_{\sigma c}. \tag{A2}$$

Where $I_{orig}$ is the index value at the calibration resolution and $I_{\sigma c}$ is the index value in the broadened spectrum. For all other line strengths the correction factor was calculated as in Equation A3:

$$C_i(\sigma) = I_{orig}/I_{\sigma c}. \tag{A3}$$

For each of the Lick indices measured $C_i$ versus $\sigma_c$ was plotted (Northeast 2006, figure 4.1). A 3$^{rd}$ order polynomial (Equation A4) from each plot was used to correct the NGC 2784 galaxy spectrum to the Lick resolution.

$$C_i = x_o + x_1\sigma_C + x_2\sigma_C{}^2 + x_3\sigma_C{}^3. \tag{A4}$$

The coefficients are given in Table A2 for each index. The correction factors were applied to index measurements in the BANSI software.

Published Lick star data differ from the NTT data for the same stars, since the Lick data were not flux calibrated. This difference was determined, so as to be able to correct each index to the Lick standard scale. Indices were first measured from the Lick star spectra, without the flux-calibration corrections. These measurements were compared to the indices measured by Worthey et al. (1994) and Worthey & Ottaviani (1997) for the same calibration stars. Mean index differences (dI) were calculated (Table A2). For each Lick index, the software (with appropriate corrections turned on) performs the flux calibration correction by adding the appropriate dI to the measured indices. The RMS difference between measured and published values for each index is given in the final column of Table A2, and this was used to calculate the error in the calibration to the Lick flux scale. Calibrations were qualitatively similar to those found by other authors.

| Index | $\sigma_L$ | $x_0$ | $x_1$ | $x_2$ | $x_3$ | dI | RMS |
|-------|-----------|-------|-------|-------|-------|-----|------|
| HdF | 4.64 | 0 | 2.81E-05 | -2.01E-07 | 6.69E-10 | -0.047 | 0.446 |
| CN$_1$ | 4.51 | 0 | -9.97E-08 | 2.33E-08 | -1.27E-11 | 0.010 | 0.036 |
| CN$_2$ | 4.51 | 0 | -1.43E-07 | 5.02E-08 | -1.55E-11 | 0.013 | 0.032 |
| Ca4227 | 4.34 | 1 | -1.07E-06 | 1.01E-06 | -1.23E-10 | -0.082 | 0.316 |
| G4300 | 4.17 | 1 | -6.71E-07 | 1.99E-07 | -7.84E-11 | -0.092 | 0.394 |
| H$\gamma$A | 4.04 | 0 | -7.81E-10 | 6.53E-07 | -6.64E-06 | -0.077 | 0.934 |
| H$\gamma$F | 4.04 | 0 | 3.51E-06 | -6.62E-07 | 3.66E-10 | 0.120 | 0.558 |
| Fe4383 | 3.91 | 1 | -3.29E-07 | 3.98E-07 | 4.17E-12 | 0.207 | 0.631 |
| Ca4455 | 3.87 | 1 | -1.07E-06 | 1.01E-06 | -1.23E-10 | -0.196 | 0.284 |
| Fe4531 | 3.83 | 1 | -5.19E-07 | 3.69E-07 | -8.13E-11 | 0.434 | 0.529 |
| C$_2$4668 | 3.74 | 1 | 6.43E-07 | 1.44E-07 | 1.08E-10 | -0.001 | 0.513 |
| H$\beta$ | 3.61 | 1 | -6.71E-07 | 1.99E-07 | -7.84E-11 | 0.086 | 0.170 |
| Fe5015 | 3.57 | 1 | -1.88E-06 | 6.27E-07 | -2.25E-10 | 0.105 | 0.408 |
| Mg1 | 3.57 | 0 | -3.43E-08 | 1.15E-08 | -3.95E-12 | 0.005 | 0.015 |
| Mg2 | 3.57 | 0 | -6.26E-08 | 1.21E-08 | -9.98E-12 | 0.002 | 0.011 |
| Mgb | 3.57 | 1 | 1.03E-06 | 2.72E-07 | 1.68E-10 | 0.037 | 0.176 |
| Fe5270 | 3.57 | 1 | -1.22E-06 | 5.19E-07 | -1.24E-10 | 0.006 | 0.204 |
| Fe5335 | 3.57 | 1 | 3.23E-07 | 8.79E-07 | 5.35E-11 | 0.115 | 0.290 |
| Fe5406 | 3.57 | 1 | 2.22E-06 | 7.68E-07 | 2.56E-10 | 0.080 | 0.257 |

**Table A2.** Values for calibration to Lick resolution, $\sigma_L$ (Å), given for each Lick Index. Values for $\sigma_L$ are from Table 2 of PS02. For all indices, the average offsets (dI) are in Å, except for Mg1, Mg2 and CN$_1$ and CN$_2$ (which are in magnitudes). The middle 4 columns give the unitless coefficients discussed above in relation to $C_i$. The final column is the RMS scatter between measured and published Lick star indices (in Å or magnitudes).